\documentclass[proceedings]{JHEP} 

\usepackage{epsfig}                   

\def\beq{\begin{equation}}
\def\eeq{\end{equation}}
\def\bea{\begin{eqnarray}}
\def\eea{\end{eqnarray}}
\def\bq{\begin{quote}}
\def\eq{\end{quote}}
\def\ra{\rightarrow}

\def\la{\lambda}

\def\eps{\epsilon}
\def\bq{\begin{quote}}
\def\eq{\end{quote}}
\def\ra{\rightarrow}

\def\la{\lambda}

\def\eps{\epsilon}
\def\bq{\begin{quote}}
\def\eq{\end{quote}}
\def\ra{\rightarrow}

\conference{Corfu Summer Institute on Elementary Particle Physics, 1998}

\title{Lepton Flavor Violation in SUSY models 
with \mbox{\boldmath{ $U(1)$}}--textures }

\author{{Mario E. G\'omez}\\

Physics Division, School of Technology,\\
Aristotle University of Thessaloniki, \\ 
Thessaloniki, GR-540 06, Greece\\
        E-mail: \email{mgomez@cc.uoi.gr}}

\abstract{$U(1)$ family symmetries have led to successful predictions of
the fermion mass spectrum and the mixing angles of the hadronic
sector. In the context of the supersymmetric unified theories, 
they further imply a non-trivial mass structure for the
scalar partners, giving rise to new sources of flavour violation. While 
$\tau\ra \mu\gamma$ decays are mostly expected to arise at rates  
significantly smaller than the current experimental limits, 
 the $\mu \ra e \gamma$ rare decays  impose 
important bounds on the model parameters.  
Even if universal soft-terms are assumed at the GUT scale,
when massive neutrinos are 
included in these theories, new mixings appear in the soft-terms. The 
predicted branching ratios for rare decays are in this case below the 
experimental bounds.  }
\begin{document} 


\section{Introduction} 

The SM predicts conservation of lepton flavor in the limit of zero neutrino 
masses.
In the case of massive non-degenerate neutrinos, the  amount of 
lepton flavour violation (LFV) is proportional to the factor $\Delta
m^2_\nu/M^2_W$ \cite{P..}, highly suppressing all relevant
processes. The current experimental limits are \cite{pdb}:
\bea
BR(\mu\ra e\gamma)&<& 4.9\cdot 10^{-11} \nonumber\\
BR(\tau\ra \mu \gamma)&<& 3.0\cdot 10^{-6} \nonumber\\
BR(\tau\ra e\gamma)&<& 2.7 \cdot 10^{-6}
\eea

SUSY theories assume a scalar(fermion) partner for every fermion (scalar) 
of the standard model. The new interactions introduced by these theories 
can generate LFV diagrams, as the ones in Fig.[1]. In the limit of flavor universal soft terms, leptons and
sleptons can be simultaneously diagonalized and hence LFVB processes will be 
suppressed. Since experimental limits for these processes are very 
restrictive, some flavor dependence in the Soft-Breaking structure will 
be reflected in an important LFV at low energies.

In SUSY-GUT's theories even if universal sof-terms are universal at 
$M_{Planck}$, non universalities are radiatively generated at the GUT 
scale due to the evoution of quarks and leptons in the same 
multiplets \cite{oldflav}. When a phenomenologically aceptable
{\it ansatze} about Yukawa structures is introduced in these theories, new 
sources of LFV appear due to a non-minimal Higgs sector and 
additional Symmetries \cite{j}.
 
When right-handed neutrinos enter in the model, Dirac mass matrices 
arise of the order of the up-quark masses.  Charged leptons and neutrinos 
are no longer diagonal in the same basis and a leptonic mixing 
matrix, similar to the $V_{CKM}$ one for the quarks, is unavoidable. 
    Moreover, it enters 
     in the construction of the $12\times 12$- sneutrino mass matrix
     which in principle would have the potential to give rise to 
     additional flavour-violating effects. Nevertheless, since only the
     effective light sneutrino mass matrix is relevant in the calculation,
     we will show in this work that the $m_D$ effects are canceled at
     first order, when the see-saw mechanism is applied.

In this talk we summarize the results of some recent work \cite{nos}, in 
which we analyze the
branching ratios for lepton flavor violating decays in a class of  models 
predicting fermion masses throught one simple  $U(1)$-family symmetry.

\begin{figure}
\begin{center}
\epsfig{file=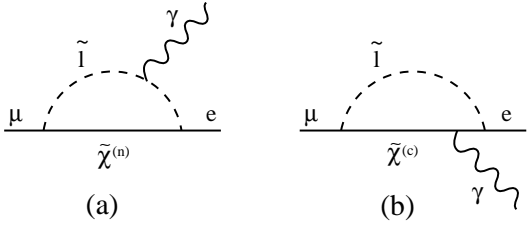,width=7cm}
\end{center}
\caption{The generic Feynman diagrams for the $\mu\ra e\gamma$
decay. $\tilde l$ stands for charged slepton (a) or sneutrino (b), while
$\tilde\chi ^{(n)}$ and $\tilde\chi ^{(c)}$ represent  
neutralinos and charginos respectively.}
\label{figure1}
\end{figure}

\section{The model}

We consider a $SU(3)\times SU(2)\times U(1)$ SUSY model with an additional 
$U(1)$-family ($U(1)_f$) symmetry as the one proposed by 
Ib\'a\~nez and Ross \cite{IR}. 
Massive right-handed neutrinos are included in the model allowing 
neutrinos to obtain a small mass via a ``see-saw'' mechanism.
The relevant superpotential terms are:
\beq
{\cal W}_{lep}= {e^c}^T\la_e\ell H + N^c \la_D\ell H' + \lambda_N \chi N^c N^c
\eeq
where $\ell$ is the left lepton doublet, $e^c$ is the right 
singlet charged 
lepton, $H$ and $H^\prime$ are  higgs doublets, $N^c$ is the  right-handed (RH)
 neutrino field  and 
$\lambda_{e,D}$ represent  Yukawa
coupling matrices in flavour space. In addition, soft supersymmetry
breaking terms generate  mass matrices for the charged slepton fields,
denoted by $\tilde m_\ell$, $\tilde m_{e_R}$.

The fermion mass hierarchy is successfully obtained
using an  additional singlet field $\theta$ and  which develops a vev an order
of magnitude below the string scale $M_U$. 
Throughout our calculations, we will further assume the existence
of a  grand unified symmetry $M_{GUT}$ without specifying the gauge group. 
 Below the unification scale, only the MSSM spectrum is assumed,
therefore the unification point will be taken at $\sim 10^{16}$ GeV.
 When the $U(1)_f$ symmetry is exact, only the third generation
has a Yukawa term in the superpotential, whilst all mixing angles are     
zero and lighter generations remain massless and uncoupled. When $\phi$
acquires a vev, the $U(1)_f$-symmetry is broken and mass terms fill in the rest
of the  mass matrix entries with Yukawa terms suppressed by powers of 
the ratio $\phi/M_U$.

The fermion mass matrix for charged leptons predicted in this theory is:
\begin{equation}
\label{eq:lmass}
{m_\ell}= \left (
\begin{array}{ccc}
c_{11}\tilde{\epsilon}^{2|a +b|} &
       c_{12} \tilde{\epsilon}^{|a|} &
             c_{13} \tilde{\epsilon}^{|a +b|}
\\
c_{21}\tilde{\epsilon}^{|a|} &
        c_{22}\tilde{\epsilon}^{2|b|} &
          c_{23}\tilde{\epsilon}^{|b|}
\\
c_{31}\tilde{\epsilon}^{|a +b|} &
       c_{32}\tilde{\epsilon}^{|b|} &1
\end{array}
\right){m_{\tau}}
\label{ml0}
\end{equation}

where the parameter $\tilde\epsilon$ is some power of the singlet vev scaled
by the unification mass, while $a,b$ are certain combinations of the
lepton and quark $U(1)_f$-charges. The parameters $c_{ij}$ in front of the
various entries (not calculable in this simple model) are assumed to
reproduce the fermion mass relations after renormalization
group running. These parameters are usually left unspecified, here however
their exact values are necessary for a reliable calculation of the 
lepton violating processes. A successful lepton mass hierarchy 
in this case is obtained
for the choice $a = 3, b = 1$ and $\tilde\eps = 0.23$. In this case, 
a possible choice  is given by
$c_{12} = c_{21} = 0.4, c_{22} = 2.2$, with the 
rest of the coefficients being unity.

The scalar mass matrices of this model are also affected by the 
$U(1)_f$-symmetry \cite{LT}. In particular, for the sleptons we obtain at 
the GUT scale
\[
\tilde m^2_{\ell,e_R}\approx
\left (
\begin{array}{ccc}

{1} & \tilde\eps^{\mid a + 6 b\mid  }
&\tilde\eps^{\mid a + b\mid} \\
\tilde\eps^{\mid a + 6 b \mid } & {1} &
\tilde\eps^{\mid b \mid}\\
\tilde\eps^{\mid a + b\mid } & \tilde\eps^{\mid b\mid} & 1
\end{array}
\right)m_{3/2}^2 \nonumber \\
\]
\beq
\label{eq:soft}
= m^2_{3/2} I +\Delta
\eeq

The Dirac mass matrix in the above model has a similar structure. 
Due to the simple $U(1)$ structure of the theory, the powers
appearing in its entries are the same as the lepton mass matrix,
however, the expansion parameter is in general different\cite{DLLRS}.
Thus, its form is given by
\begin{equation}
\label{eq:dmass}
{m_{\nu_D}}\approx \left (
\begin{array}{ccc}
{\epsilon}^{2|a + b|} &
        {\epsilon}^{|a|} &
              {\epsilon}^{|a + b|}
\\
{\epsilon}^{|a|} &
        {\epsilon}^{2|b|}&
             {\epsilon}^{ |b| }
\\
{\epsilon}^{|a + b|} &
       {\epsilon}^{|b|}&1
\end{array}
\right){m_{top}}
\label{ml0b}
\end{equation}
The choice of charges a=3, b=1 allows to identify the Dirac mass matrix 
with the up-quark mass matrix.        
A choice of coefficients leading to correct up-quark masses
is obtained for  $c_{12}=c_{21}=.5, c_{32}=c_{23}=1.5$, with the
rest of the coefficients being unity and {\bf $\epsilon = .053$.}

\section{Sneutrino mass matrix}

The s-neutrino mass matrix is a $12\times 12$ matrix, its structure is given 
in terms of the $3\times 3$ Dirac,  Majorana  and slepton mass matrices.
 In the absence of scalar mixing effects at
the unification scale where any other source of flavour violation
is rather irrelevant in the calculation of the branching ratios,
it is generally expected that --as in the case of charged sleptons--
 the Dirac term induces considerable mixing effects. We will
show here that this is {\it not} the case in the sneutrino mass matrix. 

This $12\times 12 $ matrix 
is rather complicated and not easy to handle.
Vastly different scales are involved and numerical investigations
should be carried out with great care.
Its form is as follows:

\bea
\label{eq:tw}
\begin{array}{ c| c | c }
&  \tilde{\nu} \ \  \tilde{\nu}^{\ast}  &  \tilde{N}^c \ \  {\tilde{N^c}}^\ast\\

\hline
\tilde{\nu}^{\ast}   &&\\
&{M^2_{\tilde{\nu}}}_{LL} & {M^2_{\tilde{\nu}}}_{LR}\\
\tilde{\nu}&&\\
\hline
{\tilde{N^c}}^\ast &&\\
&{M^2_{\tilde{\nu}}}_{RL} & {M^2_{\tilde{\nu}}}_{RR}\\
{\tilde{N^c}}&&\\
\end{array}
\eea

Where all entries are $6 \times 6$ matrices given by: 

\bea
{M^2_{\tilde{\nu}}}_{LL} &=&
\left(
\begin{array}{cc}
m_{\tilde{l}}^2+ m_D^{\ast} m_D^{T}  & 0 \\
0 & m_{\tilde{l}}^2+ m_D m_D^{+} 
\end{array}
\right) \nonumber \\
&& \nonumber \\
{M^2_{\tilde{\nu}}}_{LR}&=&
\left(
\begin{array}{cc}
m_D^{\ast} M^T & \left({A^{\ast}_\nu}+ \mu cot\beta\right) m_D^{\ast}\\
\left({A_{\nu}}+ \mu cot\beta\right) m_D & m_D M^+\\
\end{array}
\right) \nonumber \\
&& \nonumber\\
{M^2_{\tilde{\nu}}}_{RL}&=&
\left(
\begin{array}{cc}
M^{\ast} m_D^T & m_D^+ \left({A^{\ast}_\nu}+ \mu cot\beta\right)\\
m_D^T \left({A_{\nu}}+ \mu cot\beta\right) & M m_D^{+}
\end{array}
\right) \nonumber \\
&& \nonumber\\
{M^2_{\tilde{\nu}}}_{RR}&=&
\left(
\begin{array}{cc}
m_N^2+M^\ast M^T 
&{A_N}^\ast M^\ast\\
 +m_D^\ast m_D^+ & \\
A_N M & m_N^2+M M^+ \\
&  +m_D m_D^+ 
\end{array}
\right) \nonumber
\eea

One can construct an effective $6 \times 6$ matrix
for the light sector, by applying matrix
perturbation theory, similar to the see-saw mechanism.
The result up to second order has the form:

\bea
\left(
\begin{array}{cc}
 {({M_{\tilde{\nu}}^2}_{eff})}_{LL} &{({M_{\tilde{\nu}}^2}_{eff})}_{LR}\\
{({M_{\tilde{\nu}}^2}_{eff})}_{RL} &{({M_{\tilde{\nu}}^2}_{eff})}_{RR}\\
\end{array}
\right) 
\eea
Where all entries are $3\times 3$ matrices given by:

\bea
{({M_{\tilde{\nu}}^2}_{eff})}_{LL}&=& 
m_{\tilde{\ell}}^2 -                    
(A_\nu+\mu\cot\beta)\cdot\nonumber\\
&&(A_\nu - 2 A_N)(m_D M^{-2} m_D^\dagger) \nonumber\\ 
{({M_{\tilde{\nu}}^2}_{eff})}_{LR}&=& 
 ( (2 A_\nu + A_N) + 2 \mu \cot\beta) \cdot \nonumber\\ 
&&(m_D M^{-1} m_D^\dagger) \nonumber\\
{({M_{\tilde{\nu}}^2}_{eff})}_{RL}&=& 
( (2 A_\nu + A_N) + 2 \mu \cot\beta)\cdot \nonumber\\
&&(m_D M^{-1} m_D^\dagger)^* \nonumber\\
{({M_{\tilde{\nu}}^2}_{eff})}_{LL}&=& 
m_{\tilde{\ell}}^2 -                    
(A_\nu+\mu\cot\beta)\cdot\nonumber\\
&&(A_\nu - 2 A_N)(m_D M^{-2} m_D^\dagger)\nonumber 
\eea

The first and second order terms are obtained asuming all parameters
as real and the A-matrices proportional to the identity. 
Notice that the second order term in the $LL$ and $RR$ pieces
can be neglected. However the first order in
the $LR$ and $RR$ pieces must be retained, since they lead to
complete mixing of the pairwise degenerate states.
This however, does not affect the flavor-violating 
branching ratio.

The simplicity of this result is rather astonishing.
Moreover, there is an additional benefit, since the complication of
the initial $12\times 12$ mass matrix can now be avoided. 
A direct numerical calculation
of mass eigenstates and mixing angles
would be a hard task, due to the vastly
different scales. 

\section{Scalar mass matrices}
If we consider common scalar masses and trilinear terms at the GUT scale,
leptons and sleptons will be diagonal in the same superfield basis. However,
due to the presence of (a) the non diagonal GUT terms
$\Delta$  at the GUT scale,
and  (b) the appearance of
$\lambda_D$ in the RG equations, the  lepton Yukawa
matrix  and the slepton mass matrix  can not be brought simultaneously
to a diagonal form at the scale of the heavy Majorana masses.
Therefore, lepton number
will be violated by the one loop diagrams of Fig. 1.

We define the unitary matrices diagonalizing the Yukawa mass textures
$\lambda_D$ and $\lambda_e$, as follows
\begin{eqnarray}
\label{eq:diag}
\lambda_D^\delta&=&T_R^T \lambda_D T_L \\
\lambda_e^\delta&=&V_R^T \lambda_e V_L
\end{eqnarray}
Here, the index $\delta$ indicates
a diagonal form. Then, the mixing
matrix $K$ in the lepton sector, defined in   
analogy to $V_{CKM}$  is given by the product
\begin{equation}
K=T_L^\dagger V_L
\end{equation}
The charged slepton masses are obtained by numerical diagonalization of
the $6 \times 6 $ matrix
\begin{equation}
\label{eq:66}
\tilde{m}_e^2=\left(\begin{array}{cc} m^2_{LL}&m^2_{LR}\\
                                   m^2_{RL}&m^2_{RR} \end{array}\right)
\eeq
where all entries are $3 \times 3$ matrices in the flavour space.
In the superfield basis where $\lambda_e$ is diagonal, it is convenient 
for later use to write the $3\times 3$ entries of (\ref{eq:66}) in the form:
\begin{eqnarray}
m_{LL}^2&=& (m_{\tilde{l}}^\delta)^2+ \delta m_N^2+\Delta_L+m_l^2 \nonumber \\
&& +M_Z^2(\frac{1}{2} -sin^2\theta_W) cos 2\beta\\
m_{RL}^2&=& (A_e^\delta +\delta A_e + \mu tan\beta) m_l\\
m_{LR}^2&=& m_{RL}^{2\dagger}
\end{eqnarray}
Each component  above has a different origin and gives an  independent 
contribution in the Branching Ratios.

We further wish to emphasize the following:
\begin{itemize}
\item $ (m_{\tilde{l}}^\delta)^2, (m_{\tilde{e_R}}^\delta)^2, A_e^\delta$ 
denote the scalar diagonal contribution of the corresponding matrices;
their  entries are obtained by numerical integration of the
RG. We consider $m_{3/2}^2$ as the common
initial condition for the masses at the GUT scale, while 
the trilinear terms scale as $a m_{3/2}$.
Since in the RGEs
we consider only third generation Yukawa couplings and common
initial conditions at the GUT scale for the soft masses, our treatment is
equivalent to working in
superfield basis, such that:
(i)  $\lambda_D$ is diagonal
from the GUT scale to the intermediate scale and
(ii) $\lambda_e$ is diagonal from the intermediate scale to low energies.
The change of bases will produce a shift  in the diagonal elements of the soft
mass matrices
at the GUT and at the intermediate scale.
This effect is negligible
(less than one percent).

\item  $\delta m_N^2$ and $\delta A_{l}$ stand for  the
off-diagonal terms which appear due to
the fact that
$\lambda_D$ and $\lambda_e$ may not be diagonalised
simultaneously.
The intermediate scale 
that enters in the calculation
(which is the mass scale for the neutral Majorana
field $M_N$) is defined by demanding that neutrino masses $\approx 1 eV$
are generated via the
``see-saw'' mechanism. This sets the $M_N$ scale 
to be around the  value $10^{13}$ GeV.
Then, the following values are obtained:
\begin{eqnarray}
\delta m_N^2&=& K^\dagger \left[ m_{\tilde l}^2(m_N)\right]K|_{ nondiag.}
\label{dN}\\
\delta A_l&=& V_L A_l(m_N) V_L^\dagger |_{ nondiag.}
\label{dA}
\end{eqnarray}

\item
The following values for
$\Delta_L$ and  $\Delta_R$ are defined at the GUT scale:
\begin{eqnarray}
\Delta_L&=& V_L^\dagger  \Delta V_L\\
\Delta_R&=& V_R^\dagger  \Delta V_R
\end{eqnarray}
\end{itemize}      
The effective $3  \times 3$
sneutrino mass matrix squared has the same form as the $m_{LL}^2$
part of the $6 \times 6$ charged slepton one,
with the difference that now Dirac masses
are absent (in consistency with
what we have shown in the analysis of the
$12 \times 12$ sneutrino matrix). Thus,
\begin{equation}
\tilde{m}_{\nu}^2= (m_{\tilde{l}}^\delta)^2+ \delta m_N^2+\Delta_L + \frac{1}{2}
M_Z^2 cos 2\beta
\end{equation}


\begin{figure}

\begin{center}
\hspace*{-1cm}
\epsfig{figure=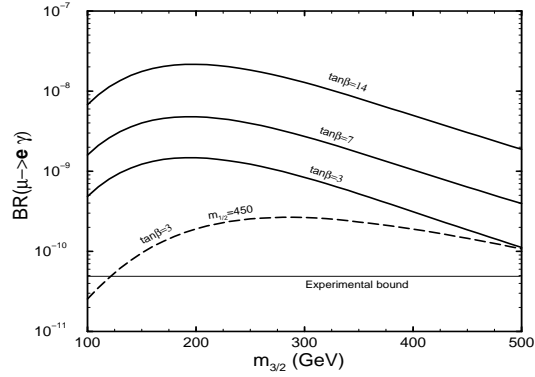,height=5truecm,width=7truecm}
\vspace{.5cm}
\caption{$BR$$(\mu\rightarrow e \gamma)$ for a range of values of
$\tan\beta$ (labeled above). Non universal  
soft masses at the GUT scale 
are considered ($\Delta \neq 0$). 
Solid lines are obtained using, $m_{1/2}=300$ GeV, and $A_0=-1.5 m_{3/2}$ as 
input parameters.}
\label{fig:ontb}
\end{center}

\end{figure}

\section{Results and Conclusions}

The branching ratio formulae for 
the $\mu\ra e\gamma$ and $\tau\ra \mu \gamma$ decays involve the masses
of  most of the supersymmetric particles. It is important therefore
for any given set of GUT parameters to know precisely all masses and 
the other low energy parameters. In the present work, 
 this is obtained by numerical integration
of the renormalization group equations of the MSSM with right handed neutrinos.
We evaluate the coupling constants, using renormalisation-group
equations at two loops.
Threshold effects are also taken
into account, by decoupling every sparticle
at the scale of its running mass $Q=m_i(Q)$.
Below  the scale $m_t$, we use the SM  beta functions..
                                              
Our analysis uses as input  
values  the unified coupling constant $\alpha_G$, at
the GUT  scale, these are determined at the meeting point of the tree 
couplings. The third generation Yukawa couplings $\lambda_{t}$, 
$\lambda_{b-\tau}$, determined by using the experimental values of 
$m_t$ and $m_\tau$ respectively. The (effective) Higgs bilinear 
coupling $\mu$ (up to its sign) is 
determined by using the 
radiative electroweak breaking condition,
and the ratio of the Higgs vev's
described by $tan\beta$ and  the flavour-symmetric
soft-breaking parameter $A_0$. We then explore the values of
the $BR$ for all significant values of the input
parameters $m_{1/2}, m_{3/2}$, $ A_0$ and $tan\beta$ and sign of $\mu$.

If we consider common scalar masses and trilinear terms at the GUT scale,
leptons and sleptons will be diagonal in the same superfield basis. However,
due to the presence of (a) the non diagonal GUT terms
$\Delta$  at the GUT scale,
and  (b) the appearance of
$\lambda_D$ in the RG equations, the  lepton Yukawa
matrix  and the slepton mass matrix  can not be brought simultaneously
to a diagonal form at the scale of the heavy Majorana masses.

\begin{figure}

\begin{center}
\hspace*{-1cm}
\epsfig{figure=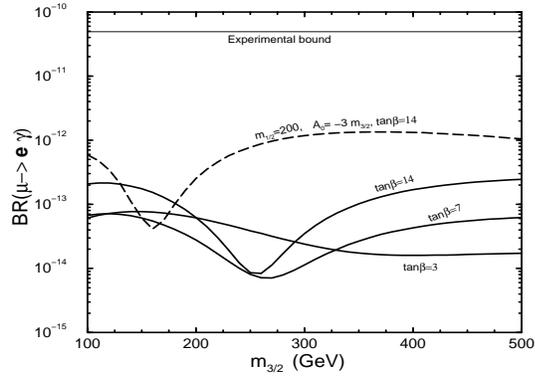,height=5truecm,width=7truecm}
\vspace{.5cm}
\caption{$BR$$(\mu\rightarrow e \gamma)$ for a range of values of
$\tan\beta$ (labeled above). Universal soft masses at the GUT scale 
are considered ($\Delta = 0$). 
Solid lines are obtained using 
 $m_{1/2}=300$ GeV and $A_0=-1.5 m_{3/2}$ as input parameters.}
\label{fig:offtb}
\end{center}

\end{figure}

In presenting our results we must consider that the mixing terms 
introduced in the
the scalar matrices by the $U(1)_f$ symmetry ($\Delta$ in 
eq.\ref{eq:soft}) are one order of 
magnitude higher than the ones due to the presence of right-handed 
neutrinos in the theory. We consider separately two distinct cases:

\begin{itemize}
\item{a)} Case with scalar mass  mixing at the GUT scale, $\Delta\neq 0$. 
Terms $\Delta_{L,R}$, are independet of $\tan\beta$ and $m_{1/2}$, since 
they are much bigger than $\delta m^2_N$ and $\delta A_e$, the 
effects of due to the presence
of right-handed neutrinos in the theory will be erased in this case.
Fig.[2] shows that parameter space is very resticted in this kind of 
models, predictions for $\mu\rightarrow e \gamma$ are inside the 
experimental limits  for low hight values
of the gaugino masses combined with low values of $\tan\beta$ and scalar masses
(dashed line). 
An alternative choice of the $U(1)_f$ charges can lead to a 
succesful prediction for the lepton masses. In ref. \cite{LT} the 
parameter b of eq. \ref{eq:lmass} is set to 1/2, however this choice
of $U(1)_f$ charges increase the size of the mixings in the scalar 
matrices leading to a $BR(\mu \rightarrow e \gamma)$ one order of 
magnitude bigger than the results presented here.

\item{b} Case without scalar mass mixing at the GUT scale, $\Delta =0$.
 We  consider the case where the scalar mass matrices
are protected from mixing effects by some kind of symmetry not
affecting the fermion mass sector. 
Our results can be summarized in Fig.[3]. Branching ratios increases 
as $\tan\beta$ and
 $A_0$ increase and decrease with $m_{1/2}$. The sign of $A_0$ has little influence 
on the final result. In the dashed line we consider the set of input parameters
leading to higher values for the branching ratio, the obtained values are two orders
of magnitude below the current experimental bounds.
\end{itemize}
We can summarize our results:
A simple $U(1)$ symetry can explain the hierarchy of the fermion masses to 
a good approximation. However when this symmetry is implemented to the 
scalar sector the amount of LFV introduced by the theory exceeds to 
experimental limits for most of the parameter space.

Assuming an additional symmetry, such that lepton scalar masses remain 
universal at the GUT scale, LFV arises from the presence of right-handed 
neutrino masses in the theory. The full effects of the $12\times 12$ 
sneutrino matrix has been taken into account. In this case the calculated 
branching ratios are below the experimental limits, but still 
interesting for a future improvement in the experimental bounds.

{\bf{Acknowledgments}}

This work is supported by E.U. under the TMR contract 
``Beyond the Standard Model'', 
ERBFMRX-CT96-0090.

\end{document}